\documentclass[prx,twocolumn,amsmath,amssymb,superscriptaddress,longbibliography]{revtex4-2}

\usepackage{color,times}
\usepackage{dcolumn}
\usepackage{multirow}
\usepackage{amssymb,amsfonts,amsmath,graphicx}
\usepackage{epsfig}
\usepackage{xcolor}
\usepackage{SIunits}
\usepackage{braket}
\usepackage{epstopdf}
\usepackage[utf8]{inputenc}
\usepackage[english]{babel}
\definecolor{darkblue}{rgb}{0, 0, 0.8}
\usepackage[colorlinks=true, breaklinks=true, linkcolor=darkblue, citecolor=darkblue, urlcolor=darkblue]{hyperref}

\begin{document}
	
\title{A non-equilibrium superradiant phase transition in free space}
	
\author{Giovanni Ferioli}
\email{giovanni.ferioli@institutoptique.fr}
\affiliation{Universit\'e Paris-Saclay, Institut d'Optique Graduate School, CNRS, 
		Laboratoire Charles Fabry, 91127, Palaiseau, France}
\author{Antoine Glicenstein}
\affiliation{Universit\'e Paris-Saclay, Institut d'Optique Graduate School, CNRS, 
		Laboratoire Charles Fabry, 91127, Palaiseau, France}
\author{Igor Ferrier-Barbut}
\affiliation{Universit\'e Paris-Saclay, Institut d'Optique Graduate School, CNRS, 
		Laboratoire Charles Fabry, 91127, Palaiseau, France}
\author{Antoine Browaeys}
\affiliation{Universit\'e Paris-Saclay, Institut d'Optique Graduate School, CNRS, 
		Laboratoire Charles Fabry, 91127, Palaiseau, France}

\begin{abstract}
A class of systems exists in which dissipation, external drive and interactions compete and give 
rise to non-equilibrium phases that would not exist without the drive. 
There, phase transitions could occur without the breaking of any symmetry, 
yet with a local order parameter -- in contrast with the Landau theory of phase transitions at equilibrium. 
One of the simplest driven-dissipative quantum systems consists of two-level atoms enclosed 
in a volume smaller than the wavelength of the atomic transition cubed, driven by a light field. 
The competition between collective coupling of the atoms to the driving field and their cooperative 
decay should lead to a transition between a phase where all the atomic dipoles are phase-locked 
and a phase governed by superradiant spontaneous emission. Here, we realize this model using a 
pencil-shaped cloud of laser-cooled atoms in free space, optically excited along its main axis, 
and observe the predicted phases. Our demonstration is promising in view of obtaining free-space 
superradiant lasers or to observe new types of time crystals.
\end{abstract}

\maketitle

Systems of  interacting particles at equilibrium exhibit 
collective phenomena such as the existence 
of phases and transitions between them. They result from an interplay between 
interactions and the action of external parameters.
Equilibrium properties of quantum many-body systems are intensively studied 
in particular in a quantum simulation approach~\cite{GeorgescuRMP2014}. 
Here instead, we experimentally investigate a model consisting of a collection of atoms driven by a resonant laser field,
where  two non-equilibrium phases now result from a competition between the drive and the dissipation 
(e.g. \cite{parmee2018phases, olmos2014steady, parmee2020,muniz2020exploring}). 
A key feature of the model, which we will refer to here as the driven Dicke model (DDM),
is the identical (cooperative) coupling of all the atoms to the electromagnetic field, 
a fact automatically ensured in sub-wavelength samples
\cite{dicke1954,agarwal1977collective,narducci1978transient,Carmichael_1977,walls1978non,Walls_1980,Hannukainen2018}. However, confining an ensemble of emitters 
in a sub-wavelength volume is experimentally very challenging in the optical regime. 
For extended ensembles, the condition of cooperative coupling to the field has thus been realized 
by placing the emitters in a cavity where they share the same electromagnetic mode \cite{ritsch2013cold}. 
Superradiant phase transition \cite{baumann2010dicke,Klinder2015} as well as dissipative time crystals \cite{Iemini2018,Kessler2021}
have been observed in this system.
Superradiant lasing has also been obtained \cite{meiser2009prospects,bohnet2012steady, norcia2016cold,laske2019,schaffer2020lasing}, but 
finding cavity-free configurations sustaining steady-state superradiance 
could simplify experiments and be of interest in  metrology.

In this work, we realize the DDM in {\it free space}, using 
a pencil-shape cloud of up to $N\approx 2000$ cold atoms, optically excited along its main axis. 
By measuring both the atomic and photonic degrees of freedom, 
we characterize  the two non-equilibrium phases predicted  by the model. 
In particular, we observe the characteristic
$N^2$ scaling  of the photon emission rate in the superradiant phase, 
thus demonstrating steady-state superradiance in free space. Finally, we observe a modification of the 
statistics of the superradiant light as we cross the phase transition. 

\begin{figure*}[!]
	\includegraphics[width=\linewidth]{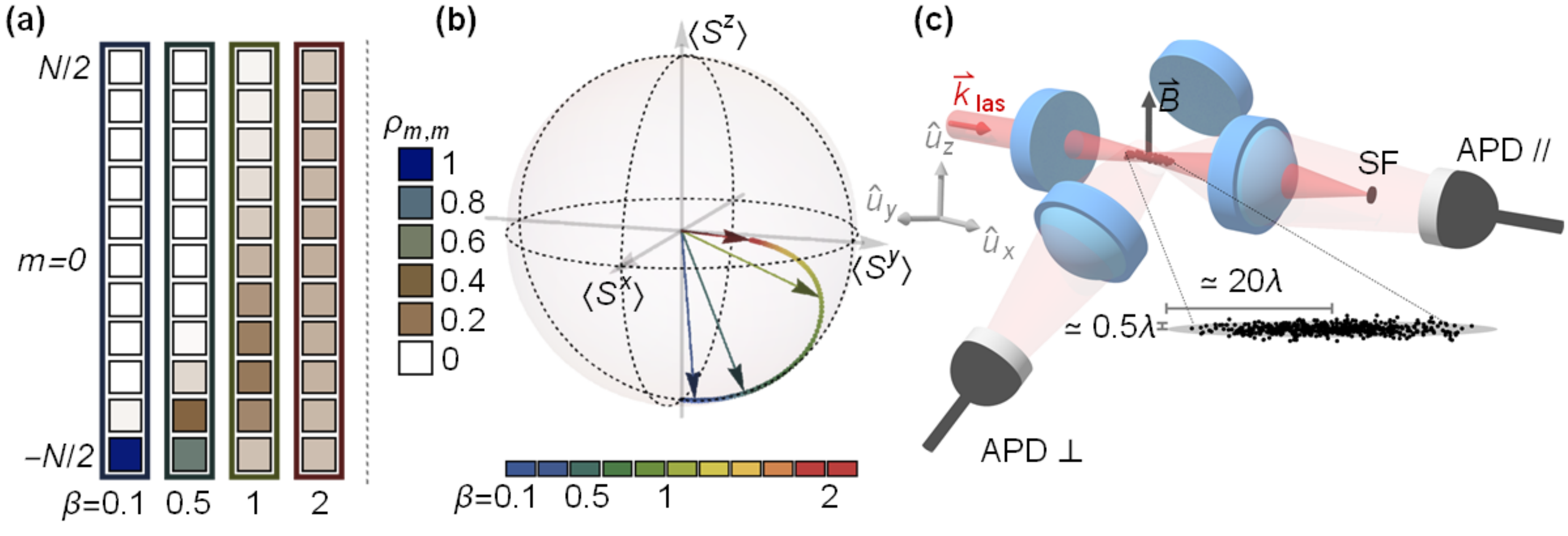}
	\caption{\textbf{Non-equilibrium phases in the Driven Dicke Model.} 
	(a) Populations of the $N+1$ states in the Dicke ladder ($N=10$), corresponding to the vectors reported in (b).
	(b) Bloch-sphere representation of the collective spin predicted by the steady-state DDM, for different values of $\beta$. 
	The collective dipole is $\langle \hat S^-\rangle = -i\langle \hat Sy\rangle$.
 	(c) Experimental setup. A pencil-shape cloud of laser-cooled $^{87}$Rb atoms is prepared in a dipole trap (not shown),
	placed between four high-numerical aperture lenses. 
	A resonant laser beam propagates along the main axis of the cloud ($1/e^2$-radius of 5\,$\mu$m). 
	Its linear polarization is perpendicular to the magnetic field $\vec B$, so that only the $\sigma^+$
	component of the light drives the atoms.
	The emitted light is collected in two different directions by two fiber-coupled avalanche photodiodes,  APD// and APD$\perp$, 
	operating in single photon counting modes. APD$\perp$ gives access to the atomic excited state population (magnetization). 
	A spatial filtering (SF) separates the laser light from the one emitted axially by the cloud, so that  APD// measures 
	the rate of superradiant light emission $\gamma_{\rm SR}(t)$.}
	\label{fig:expsetup}
\end{figure*}

The DDM describes an ensemble of $N$ two-level atoms (states $|g\rangle$ and $|e\rangle$)
as a collective spin $\hat S^{\pm}=\sum_{i=1}^N \hat\sigma_i^{\pm}$ 
(here $\hat\sigma_i^-=\ket{e_i}\bra{g_i}=(\hat\sigma_i^+)^{\dagger}$), see Methods~\ref{App:DrivenDicke}. 
The indiscernability of the atoms with respect to the field restricts the accessible states
to the permutationally  symmetric ones, $\ket{S=N/2, m=-S,..., S}$, which form a ladder
(see Fig.\ref{fig:expsetup}a). 
The Hamiltonian describing the interaction of this collective 
spin with a classical light field, resonant with the single-atom transition, is  
$\hat H_{\rm L}= (\hbar\Omega/2)(\hat S^+ +\hat S^-)$, with $\Omega$ the Rabi frequency.
The dynamics of the collective spin is governed by the equation:
\begin{equation}\label{Eq:ddMasterMT}
\frac{d\rho}{dt}=-\frac{i}{\hbar}[\hat H_{\rm L}, \rho]+\frac{\Gamma}{2} (2\hat S^-\rho \hat S^+ 
- \hat S^+\hat S^-\rho -\rho \hat S^+\hat S^-)\ ,
\end{equation}
where the last term describes the collective spontaneous emission (here $\Gamma$ 
is the single-atom decay rate from $|e\rangle$). 
In steady-state, this model supports two
non-equilibrium phases, depending on the ratio between the drive and the 
collective dissipation $\beta= 2\Omega/(N\Gamma)$ \cite{Hannukainen2018} (see Methods~\ref{App:PredictionDDM}). 
For $\beta < 1$, the atomic dipoles phase-lock and the ensemble develops a collective dipole 
$\langle \hat S^-\rangle_{\rm st} =- i\Omega/\Gamma$, as represented in Fig.\,\ref{fig:expsetup}(b). 
As $\beta$ increases, so does the amplitude of the dipole 
until it reaches its largest value $N/2$ for $\beta =1$. Conversely, the total magnetization 
$\langle \hat S_z\rangle_{\rm st}$ decreases to 0.
For $\beta \gg 1$, all the states of the ladder are equally populated (see Fig.\,\ref{fig:expsetup}a), the collective dipole vanishes
and superradiance dominates with the characteristic $N^2$ scaling of the photon emission rate. 
We will refer to the $\beta < 1$-phase as magnetized and  
the $\beta>1$-one as superradiant.  
In the limit $N\rightarrow \infty$, the value $\beta = 1$ corresponds to the critical point of a 
second order phase-transition 
(see Method \ref{App:PT}).
These are the two phases that we observe and characterize here. 

Our experiment (see Fig.\,\ref{fig:expsetup}c) \cite{Glicenstein2021} 
relies on a cloud of  up to $\simeq 2000$ laser-cooled $^{87}$Rb atoms placed in a dipole trap.  
We isolate the two states $\ket{g}=\ket{5S_{1/2}, F=2, m_F=2}$ and
$\ket{e}=\ket{5P_{3/2}, F=3, m_F=3}$ of the D2 transition 
($\lambda=2\pi/k=780$\,nm, linewidth $\Gamma\simeq 2\pi \times 6$\,MHz) using 
a $96$\,G-magnetic field. The ensemble has axial and radial sizes $\ell_{\text{ax}}=20-25\lambda$ and 
$0.5\lambda$. The mean distance between atoms in the cloud  is
$r\sim 3/k$, so that the coherent dipole interactions can be neglected in a first approximation
(see discussion at the end of the text).
For our pencil-shape ensemble of two-level atoms, Ref.~\cite{gross1982} shows that 
Eq.~(\ref{Eq:ddMasterMT}) remains valid in a mean-field model simply replacing $\Gamma$ by $\Gamma\mu$.
The parameter 
$\mu\sim \Delta\Theta/(4\pi)$ characterizes the coupling of the extended cloud to its 
diffraction mode extending over a 
solid angle $\Delta\Theta$
\cite{gross1982,AllenEberly,superradiance2017sutherland,TANJISUZUKI2011201,ritsch2013cold} : 
$\tilde N = N\mu$ is then the effective number of atoms corresponding to 
the cooperative coupling to the diffraction mode  (Method~\ref{App:extendedDDM}).
Here $\mu\sim \lambda/(2\pi \ell_{\text{ax}})\simeq 0.003(2)$ (see Method \ref{App:Cooperativity}), 
allowing us to reach $\tilde N\sim 10$, a value sufficiently large to observe the crossover between the 
two non-equilibrium phases of the DDM. 
After optically pumping the atoms in $|g\rangle$ and switching off the trap for $\sim 500$\,ns, 
we excite the cloud with 150\,ns-long pulses of a resonant laser beam propagating along  its main axis. 
With a temperature  $\simeq 200\,\mu$K, the atoms can be considered as frozen during the excitation.  
We repeat this procedure 30 times on the same cloud and average over $\sim 2000$ clouds. 
We measure  the number of emitted photons
in two orthogonal directions with avalanche photodiodes  (APDs). 
The first one (APD$\perp$), radially aligned, is sensitive to the excited state population 
$n_e(t)$, related to the magnetization $s_z(t)=2n_e(t)-1$ (see Methods \ref{App:Cooperativity}). 
This quantity acts as an order parameter for the system.
The second one (APD//) measures the photon emission rate in the superradiant mode 
$\gamma_{\rm SR}(t)=\Gamma\langle \hat S^+ \hat S^-\rangle$~\cite{AllenEberly,ferioli2021laser}.

\begin{figure}
	\centering
	\includegraphics[width=\linewidth]{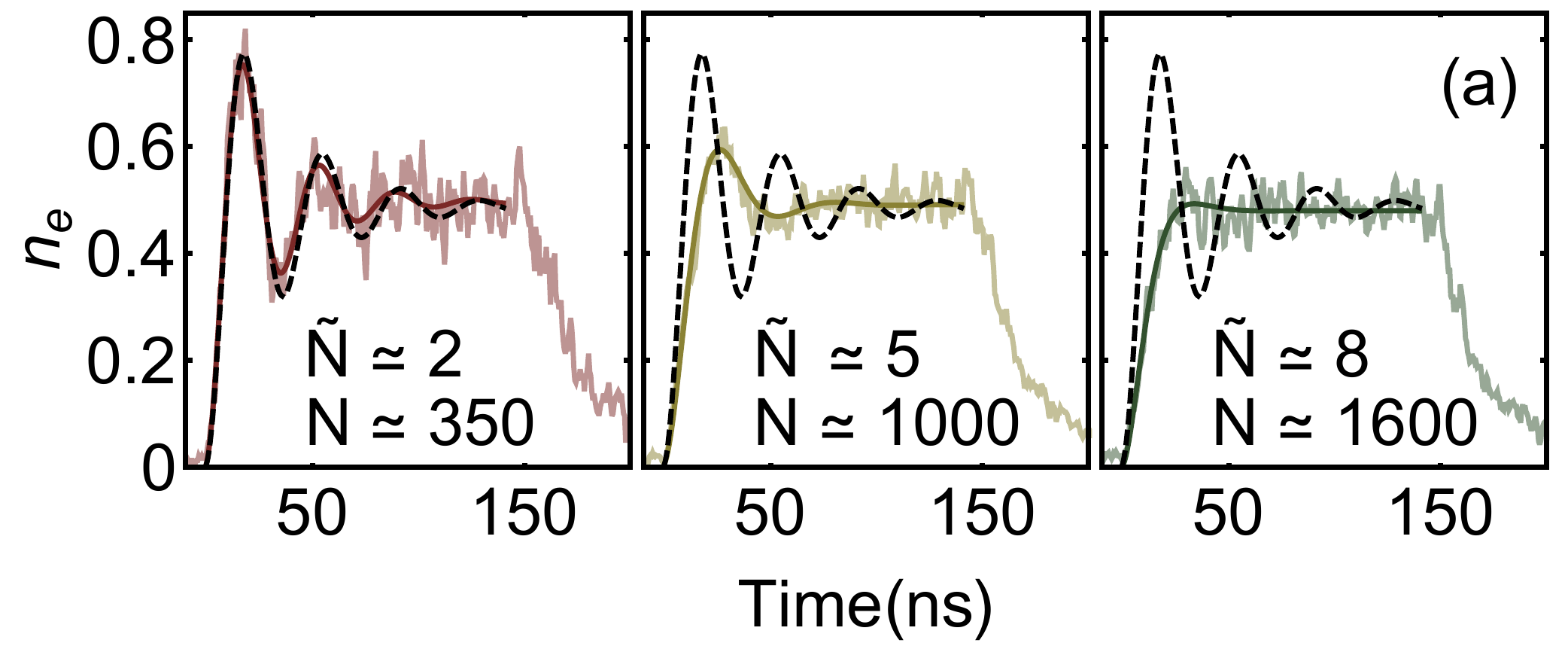}
	\includegraphics[width=\linewidth]{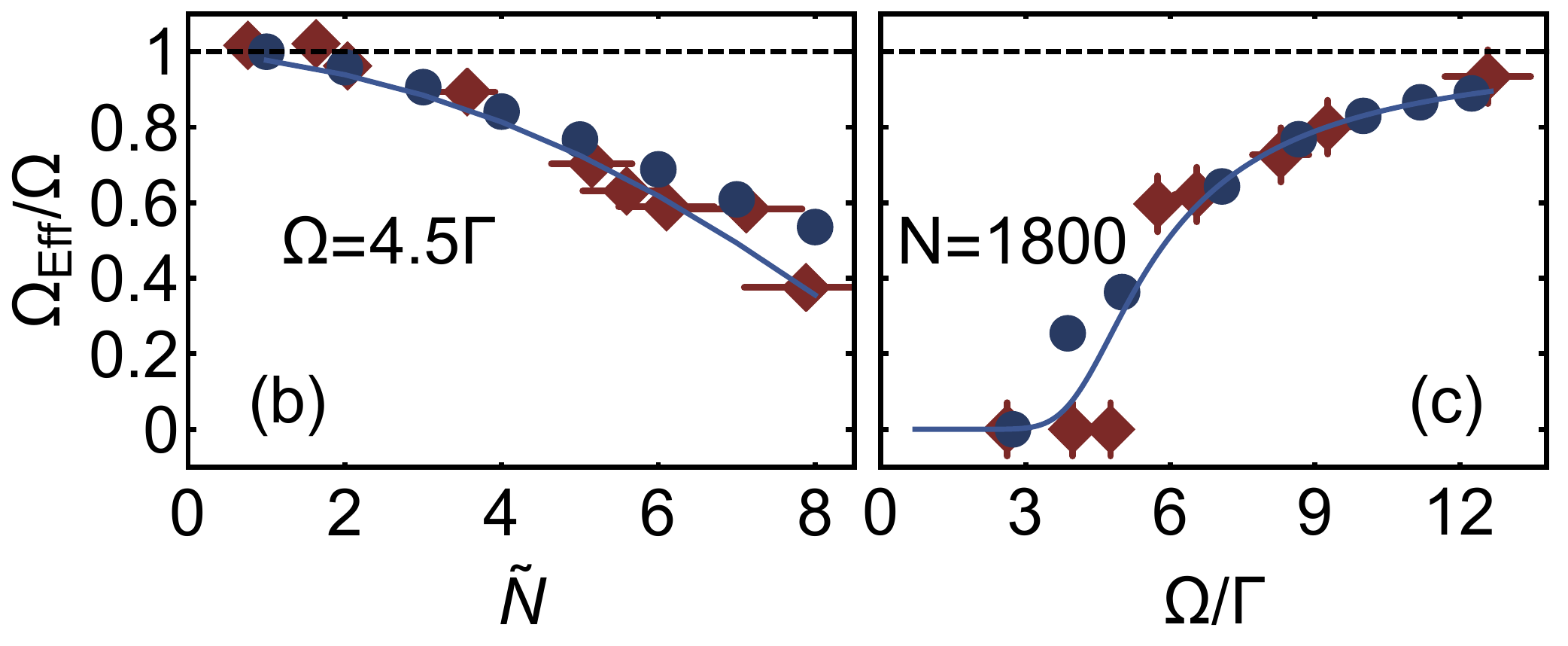}
	\caption{\textbf{Collective dynamics during excitation.} 
		(a) Excited state population $n_e(t)$ during the laser pulse measured with APD$\perp$ 
		(temporal bins: 1 ns), for different $N$. 
		Dashed black line: solution of OBEs. 
		Colored line: fit  using the analytical solution of the OBEs. 
		(b,c) Colored diamonds: experimental values $\Omega_{\text{Eff}}$ 
		as a function of $\tilde{N}=N\mu$ (b) and of $\Omega$ (c). 
		Error bars on $\Omega_{\text{Eff}}$ from the fit. 
		Uncertainties on $\tilde{N}$ and $\Omega$: 10\% shot-to-shot fluctuations,
		evaluated from a sample of 1000 repetitions.
		Blue dots: prediction from the solution of the time-dependent DDM fitted 
		as in the experiment. Continuous blue lines: $\Omega_{\text{Eff}}$
		from the steady state of the DDM. }	
	\label{fig:dynamics}
\end{figure}

\begin{figure*}
	\includegraphics[width=\linewidth]{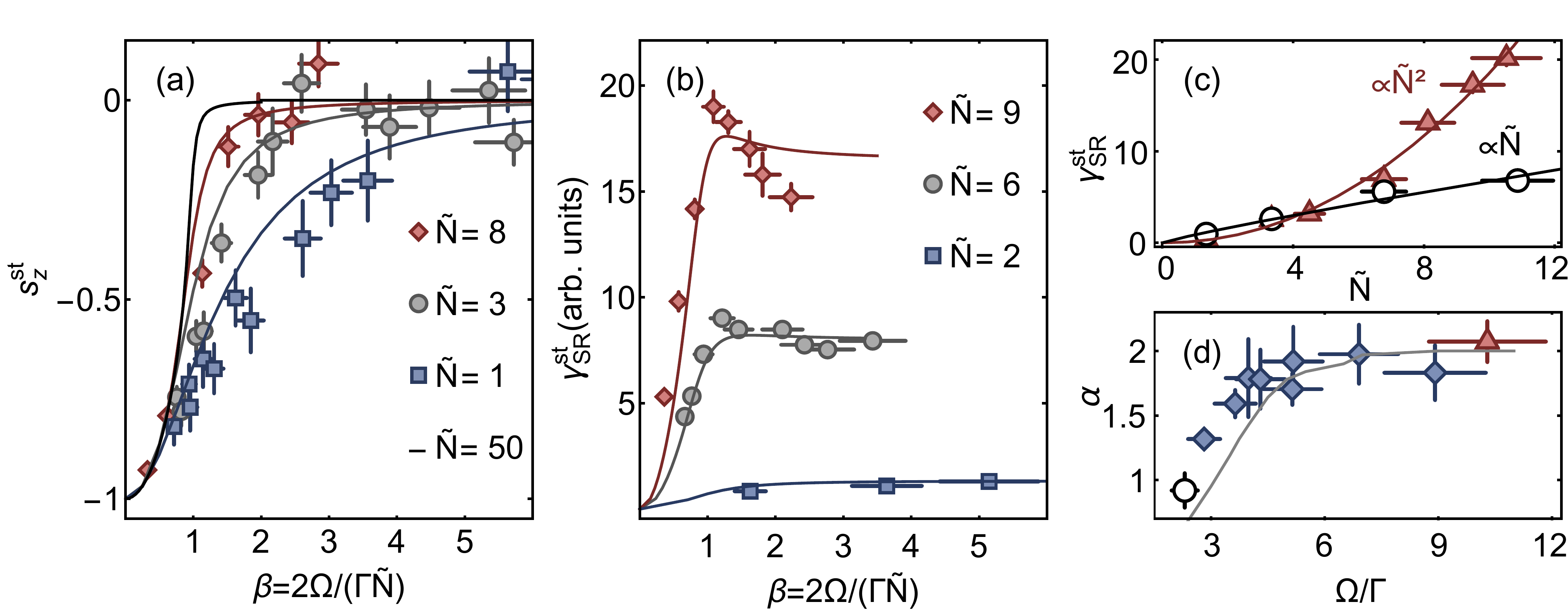}
	\caption{\textbf{Onset of the superradiant phase.} 
		(a,b) Experimental steady-state values of $s_z$ and $\gamma_{\rm SR}$ 
		as a function of $\beta$, for different atom numbers $\tilde N$. 
		Lines: predictions of the DDM.
		(c): examples of the dependence $\gamma_{\rm SR}\propto \tilde{N}^\alpha$ in the weak (black empty-circles, $\Omega\simeq2.5 \
		\Gamma$) and strong (red triangles, $\Omega\simeq 10\Gamma$) driving regimes.  
		(d) Exponent  $\alpha$ of the  fit.	Grey line: prediction of the DDM.
		Error bars on $s^{st}_{z}$ and $\gamma_{\text{SR}}^{st}$: 
		standard error on the mean from 100 repetitions of the experiment. 
		For $\alpha$: error from the fit. }
	\label{fig_steady}
\end{figure*}

We start by investigating the dynamics of the magnetization 
during the application of a laser excitation pulse. 
First, we fix the Rabi frequency of the laser driving to 
$\Omega= 4.5 \Gamma$ and vary $N$.
Examples of experimental curves for different $N$ are reported in Fig.\,\ref{fig:dynamics}\,(a).
For low $N$, the dynamics is well described by the solution of the two-level optical Bloch equations (OBEs),
indicating independent atom behavior.  As $N$ increases, we observe a reduction of 
the frequency and amplitude of the oscillations, until they vanish
for the largest $N$. We fit each curve by the analytical solution of OBEs~\cite{loudon2000quantum},  
with an effective Rabi frequency $\Omega_{\text{Eff}}$ and the decay rate as free parameters. 
Figure\,\ref{fig:dynamics}(b) reports the fitted values of $\Omega_{\text{Eff}}$, which decrease as $N$ increases.
Second, we perform the complementary experiment where we fix $N\simeq1800$ 
and vary $\Omega$. 
We observe oscillations of $n_e$ only above a critical driving strength $\Omega_{\rm c}$, and $\Omega_{\text{Eff}}$ 
becomes comparable to $\Omega$ only in the strongly driven regime ($\Omega > 10\Gamma$).

Our observations can be explained in the framework of the DDM. 
When driven by the laser, the ensemble develops a collective dipole $\langle \hat S^-\rangle$, 
which in turn radiates a field 
whose amplitude inside the cloud is $\langle \hat E_{\text{Sc}} \rangle = -i\hbar\Gamma\langle \hat S^-\rangle/d$ 
($d$ is the dipole matrix element of the $e-g$ transition)~\cite{walls1978non,Walls_1980}. 
The field $E_{\rm Eff}$ in the cloud
results from the superposition of the laser field $E_{\rm L}=\hbar \Omega/d$ and of $ \langle E_{\text{Sc}}\rangle$, 
yielding an effective Rabi frequency
$\Omega_{\text{Eff}}= d E_{\rm Eff}/\hbar=\Omega-i\Gamma\,\langle \hat S^- \rangle$.
For a resonant excitation, $\langle \hat S^- \rangle$
is purely imaginary so that $|\Omega_{\text{Eff}}|\leq \Omega$: 
the collective  dipole gives rise to a $\pi$-shifted field which screens the laser field. 
Qualitatively, the screening increases with the amplitude of the collective dipole, hence with $N$.
To compare quantitatively the data to the DDM, we solve numerically Eq.\,(\ref{Eq:ddMasterMT}) 
to get $n_e(t)$, and fit the solution with the same functional form as for the data. The only free parameter 
in the simulation is $\mu=\tilde N/N$. 
We find a good agreement between the theoretical prediction and the 
experimental results for $\mu\simeq 0.005$, as shown in Fig.\,\ref{fig:dynamics}(b,c). 
Considering the errors on the determination of the cloud sizes and atom numbers, 
this value is consistent with the inferred  one (see Methods~\ref{App:Cooperativity}).

We also calculate the {\it  steady-state} solution of Eq.\,(\ref{Eq:ddMasterMT})  
to extract $\langle \hat S^- \rangle$ and thus $\Omega_{\text{Eff}}$, using the value of $\mu$
obtained above. As visible in Fig.\,\ref{fig:dynamics}(b,c), 
the steady-state values of $\Omega_{\rm Eff}$ matches the ones extracted from the dynamics. 
This fact indicates that, for $\beta =2\Omega/\Gamma \tilde{N} <1$, the collective coherence 
giving rise to the screening is established inside 
the cloud in a timescale ($\sim1/\tilde N\Gamma$) faster than the driving period $1/\Omega$. 
Thus, $\Omega_{\text{Eff}}\approx\Omega-i\Gamma\,\langle \hat S^- \rangle_{\rm st}$. 
The existence of the threshold in $\Omega$ observed in Fig.\,\ref{fig:dynamics}(b,c) can now be understood:  
for a given $\tilde N$, and for $\beta \leq 1$ (magnetized phase), 
$\langle \hat S^- \rangle_{\rm st}=-i\Omega /\Gamma$, so that  
$\Omega_{\text{Eff}}\approx 0$ up to a critical driving strength $\Omega_{\text{c}}/\Gamma=\tilde N/2$
where the dipole reaches its largest amplitude. 
For $\Omega \gg \Omega_{\text{c}}$ ($\beta \gg 1$), the system is saturated, 
the collective dipole $\langle \hat S^- \rangle_{\rm st}$ is suppressed as 
$N/\beta$ (see Methods \ref{App:PredictionDDM_semiclas}), and $\Omega_{\text{Eff}}\simeq\Omega$.  
Conversely, for a fixed value of $\Omega/\Gamma >1$, increasing $\tilde N$ drives the system from 
the superradiant phase ($\beta>1$) where the collective dipole increases with $\tilde N$, 
hence $\Omega_{\rm Eff}$ decreases, 
to the magnetized phase ($\beta<1$) where $\Omega_{\text{Eff}}\approx 0$.

The agreement between the data and the DDM obtained 
with $\mu$ as a single free-parameter  validates the applicability of the model  for our cloud in {\it free space}
using an effective atom number $\tilde{N}$. This result allows us to 
investigate the transition between the two non-equilibrium phases predicted by the DDM.
To do so we now focus on the steady-state properties of the system.

The steady-state values of the magnetization $s_z$ (APD$\perp$) and emission rate $\gamma_{\text{SR}}$ (APD//)
are measured by averaging over a 50\,ns-time window before the end of the driving pulse. 
We report in Fig.\,\ref{fig_steady}\,(a,b) these values as a function of $\beta= 2\Omega/(\tilde N \Gamma)$ for three $\tilde{N}$, 
together with the theoretical predictions of the DDM. 
The data, plotted as a function of the scaled parameter $\beta$, show both for $s_z$ and $\gamma_{\text{SR}}$ 
a crossover between two phases. It  becomes steeper as $\tilde{N}$ increases 
and should tend towards a phase transition for $\tilde N\to\infty$ \cite{walls1978non,Walls_1980}.

To characterize further the phases, we study the dependence of $\gamma_{\text{SR}}$ with $\tilde{N}$. 
Fig.\,\ref{fig_steady}(c) presents two examples corresponding to different $\Omega$'s,
together with a polynomial fit $\gamma_{\rm SR}\propto \tilde N^\alpha$. 
As reported in Fig.~\ref{fig_steady}(d), the exponent $\alpha$ varies  from below 1 in the weak 
driving regime to 2 in the strong driving one, 	
as was also observed for superradiant lasers \cite{bohnet2012steady,schaffer2020lasing}.
Once again, this is expected from the DDM. 
For $\beta\gg1$ (superradiant phase), the populations of Dicke states are saturated and the dipole vanishes: 
superradiant spontaneous emission dominates, and $\langle \hat{S}^+ \hat{S}^-\rangle \propto \tilde N^2$. 
Conversely, in the magnetized phase ($\beta <1$) 
the system develops a collective dipole,
and $\gamma_{\rm SR}=
\langle \hat S^+\hat S^-\rangle\approx |\langle \hat S^-\rangle|^2=\Omega^2/\Gamma^2$, independent of $\tilde N$.
In the crossover between the two regimes, $\langle \hat{S}^+ \hat{S}^-\rangle \sim \tilde N$ (see Methods \ref{App:PredictionDDM_numerical}). 
The same analysis applied to the numerical solution of the DDM yields results in very good agreement with the data,
as shown in Fig.\,\ref{fig_steady}(d).     

\begin{figure}
	\includegraphics[width=0.9\linewidth]{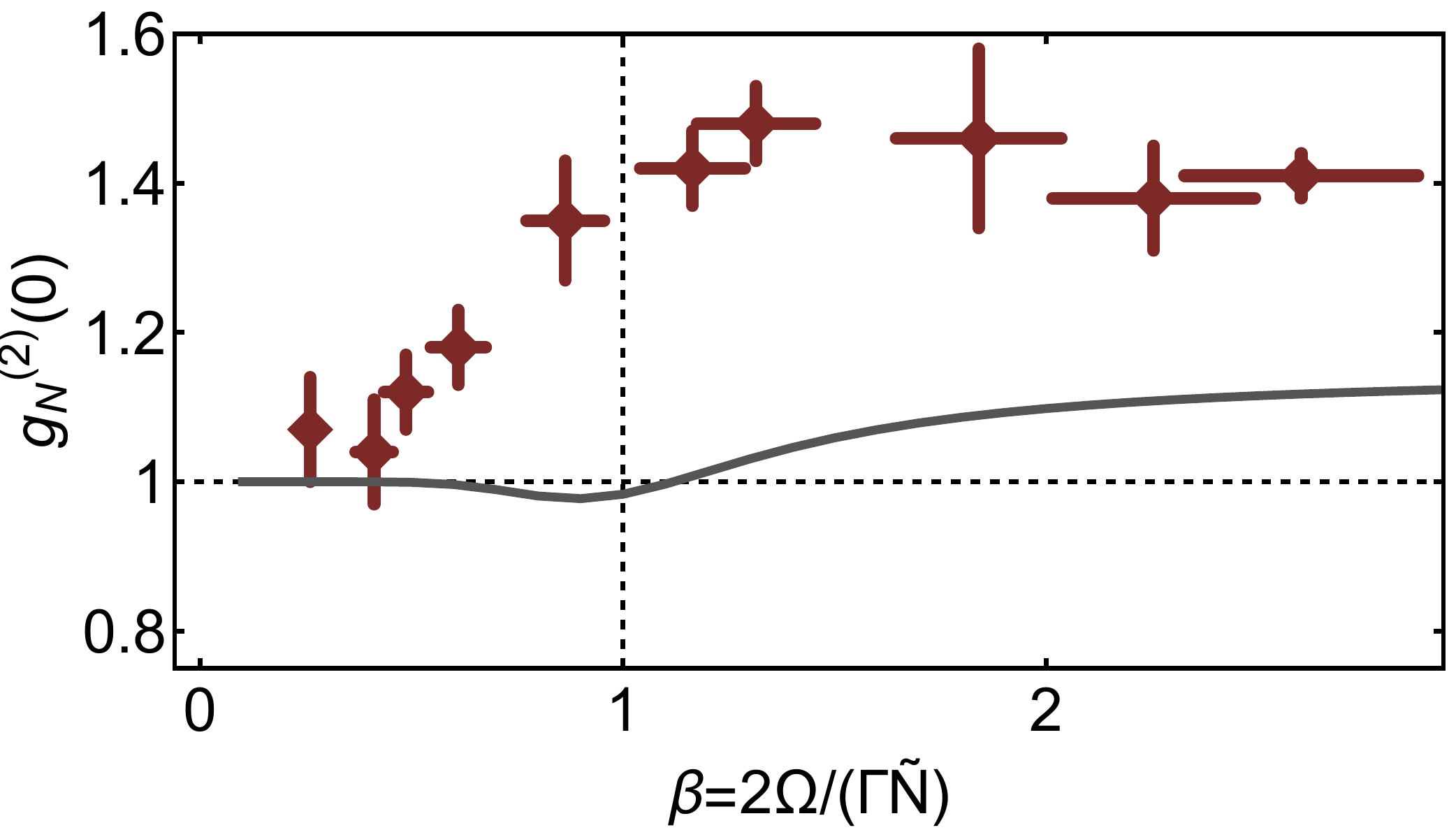}
	\caption{\textbf{Intensity correlations at equal time in the superradiant mode.} 
		Diamonds: $g_N^{(2)}(0)$ as a function of $\beta$ for $\tilde{N}\simeq 7$.
		Grey line: theoretical prediction from the DDM for $g_{N,\rm DDM}^{(2)}(0)$. 
		Error bars: standard deviation of the data  evaluated in a window of 5\,ns 
		centered around $t=0$, for 2000 repetitions of the experiment.}
	\label{fig_g2}
\end{figure}

As seen above, the transition separates a phase where a collective dipole is driven by the laser, 
from a phase where collective spontaneous emission dominates. 
We therefore expect a change in the statistics of
the light emitted by the cloud \cite{gold2022spatial, liedl2022observation} as we move through the crossover.  
To explore this, we measure 
the steady-state intensity correlations at equal times
$g_N^{(2)}(0)=\langle (\hat E_{\rm s}^-(t))^2(\hat E_{\rm s}^+(t))^2 \rangle/|\langle \hat E_{\rm s}^-(t)\hat E_{\rm s}^+(t) \rangle|^2$  
of the light field $\hat E_{\rm s}$ emitted in the superradiant mode in the far field. To do so we place a 50/50 fibered beamsplitter after 
the lens collecting this mode, and one APD in each output port \cite{loudon2000quantum}. 
We register the simultaneous coincidences in 1\,ns time bins, for the last 50\,ns of the laser pulse. 
Figure\,\ref{fig_g2} presents the measured value of $g_N^{(2)}(0)$ as a function of $\beta$. 
It does show a modification of the statistics of the emitted light around the transition:
$g_N^{(2)}(0)\simeq 1$ below threshold ($\beta \le 1$), indicating that the light  has the same 
statistics than the one of the driving laser field, as expected for a classical dipole in steady-state
(see also \cite{Shahmoon2022});  
in the strong driving regime,  $g_N^{(2)}(0)$ saturates around 1.45, thus showing bunching.
To compare to the prediction of the DDM, and as $\hat E_{\rm s}^+\propto \hat S^-$ \cite{loudon2000quantum}, we calculate 
$g_{N,\rm DDM}^{(2)}(0)=\langle (\hat S^+)^2(\hat S^-)^2\rangle/|\langle \hat S^+\hat S^-\rangle|^2$ 
\cite{hassan1980intensity,Carmichael_1980}. 
The result is presented in Fig.\,\ref{fig_g2}. 
Despite the lack of quantitative agreement with the experimental data, the DDM also predicts 
a change in the light statistics at the transition.
The quantitative mismatch between the data and the model requires further theoretical investigations. 
In particular, it may be that despite the good agreement
between the DDM and the data observed for the magnetization and $\gamma_{\rm SR}$, 
the model is too simple to calculate the statistical properties of the light emitted by an extended sample, 
as it ignores the spatial correlations between atoms. 

Finally, we compare the findings of this paper with our previous works where we investigated
dense atomic ensembles \cite{Pellegrino2014} or a chain of atoms spaced by $\sim \lambda$ \cite{Glicenstein2020}, 
driven by a near-resonant laser. In both cases, we observed in the {\it  weak driving} 
regime ($\Omega\ll\Gamma$) a shift of the line induced by the coherent dipole interactions:
for the atomic ensembles, they resulted from the large spatial density, while for the much more diluted chain they were a consequence of
constructive interferences of the fields scattered by the atoms. On the contrary, at {\it strong driving}  ($\Omega\gtrsim\Gamma$),
the shift was suppressed in the chain, owing to the fact that atomic dipoles vanishes at high intensity.  
In the present work, the atomic ensemble is about an order of magnitude more dilute than in Ref.~\cite{Pellegrino2014} and
the driving strength is such that $\Omega\geqslant 2.5\Gamma$, 
making the role of coherent dipole interactions negligible in a first approximation. 
However we do expect the dipole interactions to play a small role at very weak driving, and therefore that the DDM 
is not a faithful description of the experimental configuration for $\beta\ll 1$ (see in particular Fig.~\ref{figShift} in Methods \ref{App:extendedDDM}). 
We also show in Methods \ref{App:extendedDDM} that the coherent dipole interactions plays no role at the mean-field level. 
What is their role beyond mean-field is an important question that will be the subject of future works. 

In conclusion, we have observed the transition between a magnetized and a superradiant non-equilibrium phase
predicted by the DDM in free-space.   
The applicability of this model for an elongated sample in free-space is perhaps unexpected 
but is validated here by the good agreement between the experiment 
and the model, at least at the level of precision reached in the experiment.
Our observations raise questions that deserve further investigations. In particular, 
what is the micros\-co\-pic justification of the validity of the DDM in free-space for an extended cloud 
simply using an effective atom number \cite{Debnath2018,Zhang_2018}?  
This work also opens promising prospects for the realization of superradiant laser in free space, e.g. using 
thermal atomic beams \cite{jager2021}, or for the observation of new types of time crystals \cite{Iemini2018}. 
Finally, increasing the density of the sample could lead to the regime where 
dipole-dipole interactions between  atoms play a role and stabilize exotic non-equilibrium phases
\cite{parmee2018phases, olmos2014steady, parmee2021bistable,muniz2020exploring}.

\begin{acknowledgements}
We thank F. Robicheaux for stimulating conversations, and A.-M. Rey, J.K. Thompson, K. Moelmer,  R.T. Sutherland, J. Marino, 
Bruno Laburthe-Tolra, D. Dreon  and D. Cl\'ement
for insightful discussions. 
We thank Daniel Goncalves-Romeu, Lisa Bombieri and Darrick Chang  for insightful inputs  on the role of the coherent dipole interactions.
This project has received funding from the European Research Council 
(Advanced  grant No. 101018511, ATARAXIA)
by the Agence National de la Recherche (ANR, project
DEAR) and by the R\'egion Ile-de-France in the framework of DIM SIRTEQ (projects DSHAPE and FSTOL). 

\end{acknowledgements}

\section*{Author contributions}

GF and AG carried out the experiments and analyzed the data. 
GF, IFB and AB conducted the theoretical analysis and simulations. 
All authors contributed to the writing of the manuscript. 

\section*{Corresponding author}
giovanni.ferioli@institutoptique.fr
	
\section*{Data availability}
All data that support the plots within this paper and the Methods of this study are 
available from the corresponding author upon reasonable request.
	
\section*{Ethics Declaration}
No competing interest.

\bibliography{Superradiance_phase_transition_biblio}

\clearpage
\section*{Methods}

\subsection{The Driven-Dicke model}\label{App:DrivenDicke}

\subsubsection{Description of the model}

The Driven Dicke model considers a system of $N$-two-level atoms (resonant frequency $\omega_0$), 
all located at the same position and 
driven by a laser field with Rabi frequency $\Omega$ and detuning $\Delta$ with respect to the 
transition frequency. Since the system size is much smaller than $\lambda$, 
the state evolution is restricted to the $N+1$ permutationally symmetric states containing $m$ excitations \cite{dicke1954}. 
One thus introduces a collective spin operator $\hat{\bf S} =\sum_{i=1}^N \hat{\sigma}_i /2$ 
($\sigma_i$ are the Pauli matrices), 
and the relevant Hilbert space is spanned by the eigenstates of 
$\hat{S}_z$, $\ket{S=N/2,m}$, with $-N/2\leq m\leq N/2$. 
The actions of the collective spin operators on these states are:
\begin{eqnarray}
\begin{aligned}
	\hat{S}^2\ket{S, m}&= S(S+1)\ket{S, m}\\
	\hat S_z\ket{S, m}&= m\ket{S, m}\\
	\hat S^+\ket{S, m}&=(\hat S_x+i\hat S_y)\ket{S, m}=A_m\ket{S, m+1}\\
	\hat S^-\ket{S, m}&=(\hat S_x-i\hat S_y)\ket{S, m}=A_{m-1}\ket{S, m-1}\\
\end{aligned}
\end{eqnarray}
where $A_m=\sqrt{S(S+1)-m(m+1)}$.
The Hamiltonian describing the interaction of the collective 
spin with the light is given by 
\begin{equation}
 	\hat H_{\rm L}= \hbar\frac{\Omega}{2}\big{(} \hat S^+ +\hat S^-\big{)}- \hbar\frac{\Delta}{2}\hat S_z\ .
\end{equation}
Importantly, the coherent (spin-exchange) component of the dipole-dipole interactions between atoms 
is ignored in this simplified model: one simply assumes that it leads 
to a renormalization of the resonant frequency $\omega_0$. 

The dynamics of the system is governed by the following master equation:
\begin{equation}
\frac{d\rho}{dt}=-\frac{i}{\hbar}[\hat H_{\rm L}, \rho]+\frac{\Gamma}{2} (2\hat S^-\rho \hat S^+ 
- \hat S^+\hat S^-\rho -\rho \hat S^+\hat S^-)\ .
\label{Eq:ddMaster}
\end{equation} 
In the $|S,m\rangle$ basis, and for the resonant case $\Delta = 0$, it leads to a 
system of $N(N+1)/2$ coupled differential equations for the matrix elements 
$\rho_{m,m'}=\langle S,m|\rho|S,m'\rangle$:
\begin{equation}\label{Eq:masterDDM}
\begin{aligned}
	&\dot{\rho}_{m,m'}=-i\frac{\Omega}{2}(A_{m-1}\rho_{m-1,m'}+A_{m}\rho_{m+1,m'}\\
	&-A_{m'-1}\rho_{m,m'-1}-A_{m'}\rho_{m,m'+1})\\
	&+\frac{\Gamma}{2}(2A_{m}A_{m'}\rho_{m+1,m'+1}-A_{m-1}^2\rho_{m,m'}-A_{m'-1}^2\rho_{m,m'})\ .
\end{aligned}
\end{equation}
They can be easily solved numerically for the small atom numbers considered here. 
From the solutions, we then evaluate the expectation values of the following operators:
\begin{eqnarray}
	\begin{aligned}
		\langle \hat S_z\rangle(t)/N&=2n_e(t)-1= \frac{1}{N}\sum_{m=-S}^{S} m \rho_{m,m}(t)\\
		\langle \hat S^- \rangle(t)&=\sum_{m=-S}^{S}A_{m-1}\rho_{m,m-1}(t)\\	
		\langle \hat S^+\hat S^- \rangle(t)&=\sum_{m=-S}^{S}A^2_{m-1}\rho_{m,m}(t)\\	
		\langle \hat S^+\hat S^+\hat S^-\hat S^- \rangle(t)&=\sum_{m=-S}^{S}A^2_{m-1}A^2_{m-2}\rho_{m,m}(t)\ .	
	\end{aligned}\label{Eq:spinvariabletheo}
\end{eqnarray}

\subsubsection{Semi-classical approach}\label{App:semiclassical}
The DDM has a semi-classical limit for $N\gg 1$ 
when considering the average value of the collective spin 
$\langle \hat{\bf S}\rangle=(\langle \hat S_x\rangle,\langle \hat S_y\rangle,\langle \hat S_z\rangle)$ \cite{Hannukainen2018}. 
To see it, we use:
\begin{equation}
{d\langle \hat S_\alpha\rangle \over dt}={\rm Tr}[\hat S_\alpha {d\rho\over dt } ]
\end{equation} 
combined with the master equation (\ref{Eq:ddMaster}) and the commutation relations of the spin operators. 
This leads to the set of coupled, non-linear equations (for $\Delta =0$):
\begin{eqnarray}
	{d\langle \hat S_x\rangle \over dt}&=&  {\Gamma\over 2} 
	(\langle \hat S_x\hat S_z\rangle+\langle \hat S_z\hat S_x\rangle)-{\Gamma\over 2} \langle \hat S_x\rangle \\
	{d\langle \hat S_y\rangle \over dt}&=& \Omega\langle \hat S_z\rangle+ {\Gamma \over 2} 
	(\langle \hat S_y\hat S_z\rangle+\langle \hat S_z\hat S_y\rangle)-{\Gamma\over 2} \langle \hat S_y\rangle \\
	{d\langle \hat S_z\rangle \over dt}&=& -\Omega\langle \hat S_y\rangle-\Gamma( \langle \hat S_x^2\rangle + \langle\hat S_y^2\rangle)\ .
\end{eqnarray}

We now assume that for large spins ({\it i.e.} $N\gg 1$), 
$\langle \hat S_\alpha \hat S_\beta\rangle\approx \langle \hat S_\alpha\rangle \langle \hat S_\beta\rangle$ for $\alpha \neq \beta$. 
Neglecting the dissipative terms $\Gamma \langle \hat S_{x,y}\rangle$ of ${\cal O}(N)$ only, 
we then obtain a set of equations that conserves the total spin 
$\langle \hat{\bf S}\rangle^2=\langle \hat S_x^2\rangle+\langle \hat S_y^2\rangle+\langle \hat S_z^2\rangle=N^2/4$. 
If we now consider that $\langle \hat S_x\rangle(0)=0$, then $\langle \hat S^-\rangle(t)=-i\langle \hat S_y\rangle$ and the system 
reduces to two coupled equations:
\begin{eqnarray}
	{d\langle \hat S^-\rangle \over dt}&=& \left(i\Omega+ \Gamma \langle \hat S^-\rangle\right) \langle \hat S_z\rangle \label{Eq:semiclassicalDip} \\ 
	{d\langle \hat S_z\rangle \over dt}&=& i\Omega\langle \hat S^-\rangle - \Gamma  \left( {N^2\over 4}-\langle \hat S_z^2\rangle\right)\ .\label{Eq:semiclassicalSz}
\end{eqnarray}
The first equation shows explicitly that the dipole $\langle \hat S^-\rangle$ is driven by the effective Rabi frequency.

\subsubsection{The driven Dicke Model for a pencil-shaped cloud}\label{App:extendedDDM}

We now extend the Driven Dicke Model  to our pencil-shaped cloud,
relying on Sec.~6 of~\cite{gross1982}.
There, the dynamics of a nearly one-dimensional system (direction ${\bf k}_0$) subjected to collective 
dissipation and driving (wavevector ${\bf k}_{\rm L}$) is derived and given by Eq. (6.9):
\begin{equation}
\begin{aligned}
	\frac{d\rho}{dt}&=\frac{1}{i\hbar}[\hat H_{\rm L}+\hat H_{\rm dd}, \rho]
	-\Gamma\mu\sum_{i\ge j} (\hat{D}^+_{k_{\rm L},i}\hat{D}^-_{k_{\rm L},j}\rho\\
	&+\rho\hat{D}^+_{k_{\rm L},j}\hat{D}^-_{k_{\rm L},i}-
	\hat{D}^-_{k_{\rm L},j}\rho\hat{D}^+_{k_{\rm L},i}-\hat{D}^-_{k_{\rm L},i}\rho\hat{D}^+_{k_{\rm L},j})\ ,	
\end{aligned}
\label{eqGH0}
\end{equation}
where $\hat{D}^{\pm}_{k_{\rm L},i}=e^{\pm i {\bf k}_{\rm L}\cdot {\bf r}_i}\hat{\sigma}^{\pm}_i$. We have 
taken here ${\bf k}_0\approx {\bf k}_{\rm L}$, as experimentally relevant.
The geometrical factor $\mu$ accounts for the
coupling to the diffraction mode and is calculated in Sec.~\ref{App:Cooperativity}. 
The Hamiltonian describing the laser driving is :
\begin{equation}\label{eqGH1}	
	\hat H_{\rm L}=\frac{\hbar \Omega }{2}\sum_{i=1}^N \hat{D}^+_{k_{\rm L},i}+\hat{D}^-_{k_{\rm L},i}\ .
\end{equation}
With respect to Eq. (6.9) of \cite{gross1982}, we have included the coherent part of the resonant 
dipole-dipole interactions: 
\begin{equation}\label{eqGH2}	
 \hat{H}_{\rm dd} = \hbar \sum_{i\neq j}V_{ij}e^{i{\bf k}_{\rm L}\cdot ({\bf r}_j-{\bf r}_i)}\hat{D}^-_{k_{\rm L}, j}\hat{D}^+_{k_{\rm L}, i}
\end{equation}
where $V_{ij}$ is related to the real part of the field radiated by $i$-th atoms in the position of the $j$-th 
(see Eq.~(\ref{Eq:Esc}) of Sec.~\ref{App:field_scattered}).
This equation holds for any disordered spatial distribution.

The equation of motion for $\langle \hat D^-_{k_{\rm L}, i}\rangle$ then reads
\begin{equation}\label{Eq:total}
	\begin{aligned}
	&\frac{d\langle \hat{D}^-_{k_{\rm L}, i}\rangle}{dt}= i\Omega \langle \hat{D}^z_{k_{\rm L},i}\rangle+
	2\Gamma\mu\sum_{i\ge j}\langle \hat{D}^-_{k_{\rm L}, j} \hat{D}^z_{k_{\rm L},i} \rangle\ \\
	&-\sum_{m \neq n}V_{mn}e^{i{\bf k}_{\rm L}\cdot({\bf r_m}-{\bf r_n})}\times\\
	&(\langle \hat{D}^-_{k_{\rm L}, l}\hat{D}^+_{k_{\rm L}, m}\hat{D}^-_{k_{\rm L}, n} \rangle - 
	\langle \hat{D}^-_{k_{\rm L}, n}\hat{D}^+_{k_{\rm L}, m}\hat{D}^-_{k_{\rm L}, l} \rangle)
	\end{aligned}
\end{equation}
Now considering a one dimensional model, 
we assume that the amplitude of the dipole of a given atom $i$ depends on its position 
only via a phase factor (spin-wave approximation), i.e., $\langle\hat{D}^{\pm}_{k_{\rm L},i}\rangle=\langle \hat{D}^{\pm} \rangle$ 
(or $\langle \hat{\sigma}_i^{-}\rangle=\langle \hat{\sigma}^- \rangle e^{i{\bf k}_{\rm L}\cdot {\bf r}_i}$). 
The equation of motion obtained in this way is identical to Eq. (6.27) of \cite{gross1982} with an external resonant driving. 
Importantly, performing the mean-field approximation (e.g. 
$\langle \hat{D}^-_{k_{\rm L}, l}\hat{D}^+_{k_{\rm L}, m}\hat{D}^-_{k_{\rm L}, n} \rangle=
\langle \hat{D}^-_{k_{\rm L}, l}\rangle\langle\hat{D}^+_{k_{\rm L}, m}\rangle\langle\hat{D}^-_{k_{\rm L}, n} \rangle$), 
the contribution of the coherent part of the dipole interactions vanishes. Using also
$\langle\hat{D}^{-}\hat{D}^{z}\rangle=\langle\hat{D}^{-}\rangle\langle\hat{D}^{z}\rangle$, 
Eq.\,(\ref{Eq:total}) leads to
\begin{equation}
	\frac{d\langle \hat{D}^-\rangle}{dt}= (i\Omega + \mu N \Gamma \langle \hat{D}^-\rangle)\langle \hat{D}^z\rangle\ 
	\label{eqGH3}	
\end{equation}
It is equivalent to our Eq.\,(\ref{Eq:semiclassicalDip}), replacing 
$\langle \hat S^-\rangle = N\langle \hat\sigma^-\rangle$ 
by $\langle \hat{D}^-\rangle=\mu N \langle \hat\sigma^-\rangle$. 
This justifies the use of an effective atom number  $\tilde N=N\mu$ to describe a pencil-shape cloud.

\begin{figure}
	\includegraphics[width=0.9\linewidth]{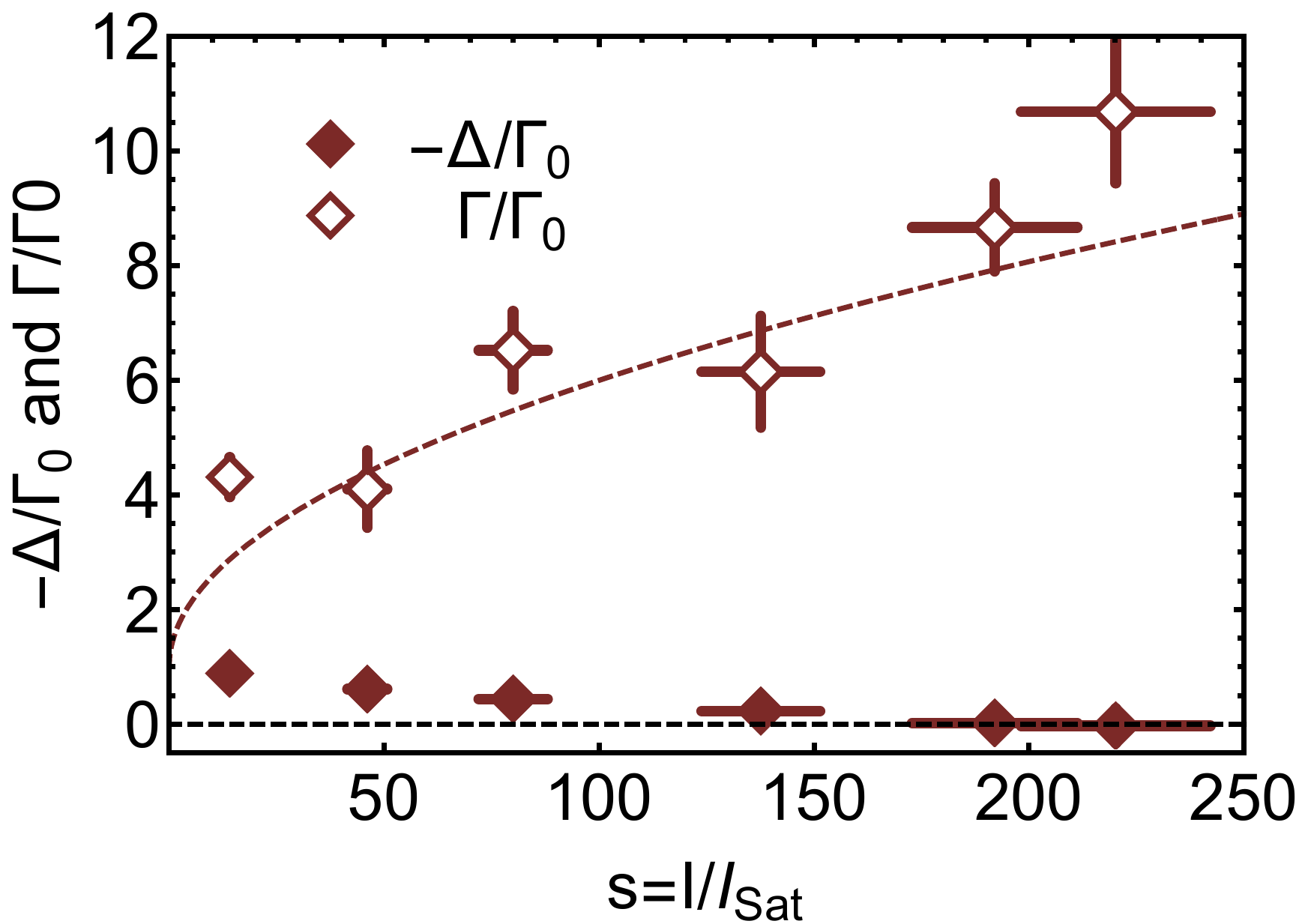}
	\caption{\textbf{Shift and width of the atomic transition} as a function of the saturation parameter
	for $N\approx 3000$. 
	Color (white) filled diamonds: shift (width). 
	The dashed line is the single atom power broadening $\sqrt{1+s/4}$.
	Error bars on $\Delta$ and $\Gamma$ are from the fit. 
	Error bars on $s=I/I_{\rm sat}$ correspond to 10\% shot-to-shot fluctuations evaluated from 1000 repetitions. }
	\label{figShift}
\end{figure}

The discussion above suggests that the Driven-Dicke model can be applied to an extended ensemble of 
randomly distributed emitters if the spin-wave and mean-field approximations are fulfilled. 
We expect this condition to be valid for a sufficiently strong driving. 
Conversely, in the weakly driven regime, this approximation is not valid and in the system 
is subject to the effect of $\hat{H}_{\rm dd}$ which manifests
as a frequency shift and broadening
of the atomic transition. To verify how appropriate the spin-wave approximation is, 
we report in Fig.\,\ref{figShift} the measured shift $\Delta$ 
and width $\Gamma$ of the  transition
in the same conditions as in the experiment, but using $\simeq 3000$ atoms. 
We observe a broadening consistent with the single-atom power 
broadening and a shift $\Delta <\Gamma_0$ for $s\ge20$. 
As the atom number is here nearly two times the largest one used in the 
experiment ($\tilde{N}\simeq10$), it means that for $\beta\ge0.3 $ the spin wave approximation can be applied. 
For smaller values of $\beta$, the applicability of the DDM has to be tested by comparing the data 
to ab-initio calculations including interactions, as the one performed in \cite{ferioli2021laser}. 

\subsubsection{Field scattered by the collective dipole inside the cloud}\label{App:field_scattered}
We used in the main text the field radiated 
inside the cloud by the collective dipole $\langle \hat E_{\text{Sc}} \rangle = -i\hbar\Gamma\langle \hat S^-\rangle/d$.
This expression corresponds to the limit $r\rightarrow 0$ (Dicke limit) 
of the imaginary part of the field radiated by a dipole $D=d \langle \hat S^-\rangle$:
\begin{equation}\label{Eq:Esc}
E_{\rm Sc} = {D\over 4\pi\epsilon_0} \left[\left({1\over r^3}-{ik\over r^2}\right)(3\cos^2\theta-1)+{k^2\sin^2\theta\over r}
\right] \, e^{ikr}\ , 
\end{equation}
using $\hbar \Gamma = d^2k^3/(3\pi \epsilon_0)$, with $k=\omega_0/c$.
The real part of $E_{\rm Sc}$ gives rise to the coherent part of the dipole-dipole interaction. 
It diverges for $r\rightarrow 0$ and is assumed to lead to 
a renormalization of the resonance frequency $\omega_0$. 
For the extended sample considered in the experiment, 
we also neglect the real part of $E_{\rm Sc}$: it would lead to a shift $\Delta\omega$ of the transition frequency 
in the low excitation intensity limit, and as here the mean interatomic distance fulfills $kr\sim 3$, we expect $\Delta\omega\lesssim \Gamma$. 
Moreover, for the Rabi frequencies used in the experiment, the shift is suppressed further~\cite{Glicenstein2020}.

\subsection{Predictions of the model}\label{App:PredictionDDM}

In order to get an intuition about the phases predicted by the DDM, 
we present here analytical and numerical, steady-state solutions of Eq.~(\ref{Eq:masterDDM}) for the range of 
parameters $\Omega/\Gamma$ and $\tilde N$ accessible in our experiment. 

\subsubsection{Steady-state solution of the semi-classical approach}\label{App:PredictionDDM_semiclas}
Equations (\ref{Eq:semiclassicalDip}) and (\ref{Eq:semiclassicalSz}) 
predict that either $\langle \hat S^-\rangle=-i\Omega/\Gamma$, in which case 
$\langle \hat S_z\rangle= (N/2)\sqrt{1-\beta^2}$ with $\beta = 2\Omega/(N\Gamma)<1$, or
$\langle \hat S_z\rangle = 0 $ and $\langle \hat S^-\rangle = -i N/(2\beta)$, for $\beta>1$. The value $\beta = 1$ thus
appears as a critical point separating two regimes: (i) for $\beta <1$, the collective spin vector lies on the $N$-atom 
Bloch sphere of radius $N/2$, and rotates around the $x$-axis, 
from the $z$-axis to the equatorial plane ($y$-axis) as $\beta$ increases up to 1.
The angle $\theta$ between the spin $\langle \hat{\bf S}\rangle $ and the $z$-axis is such that $\tan\theta = 
|\langle \hat S^-\rangle|/|\langle \hat S_z\rangle|=\beta/\sqrt{1-\beta^2}$;
(ii) when $\beta\ge 1$, the component $\langle \hat S_z\rangle$ is locked to $0$, while the component 
$\langle \hat S^-\rangle$ along the $y$-axis decreases as $\beta$ increases. 

\begin{figure}
	\includegraphics[width=0.9\linewidth]{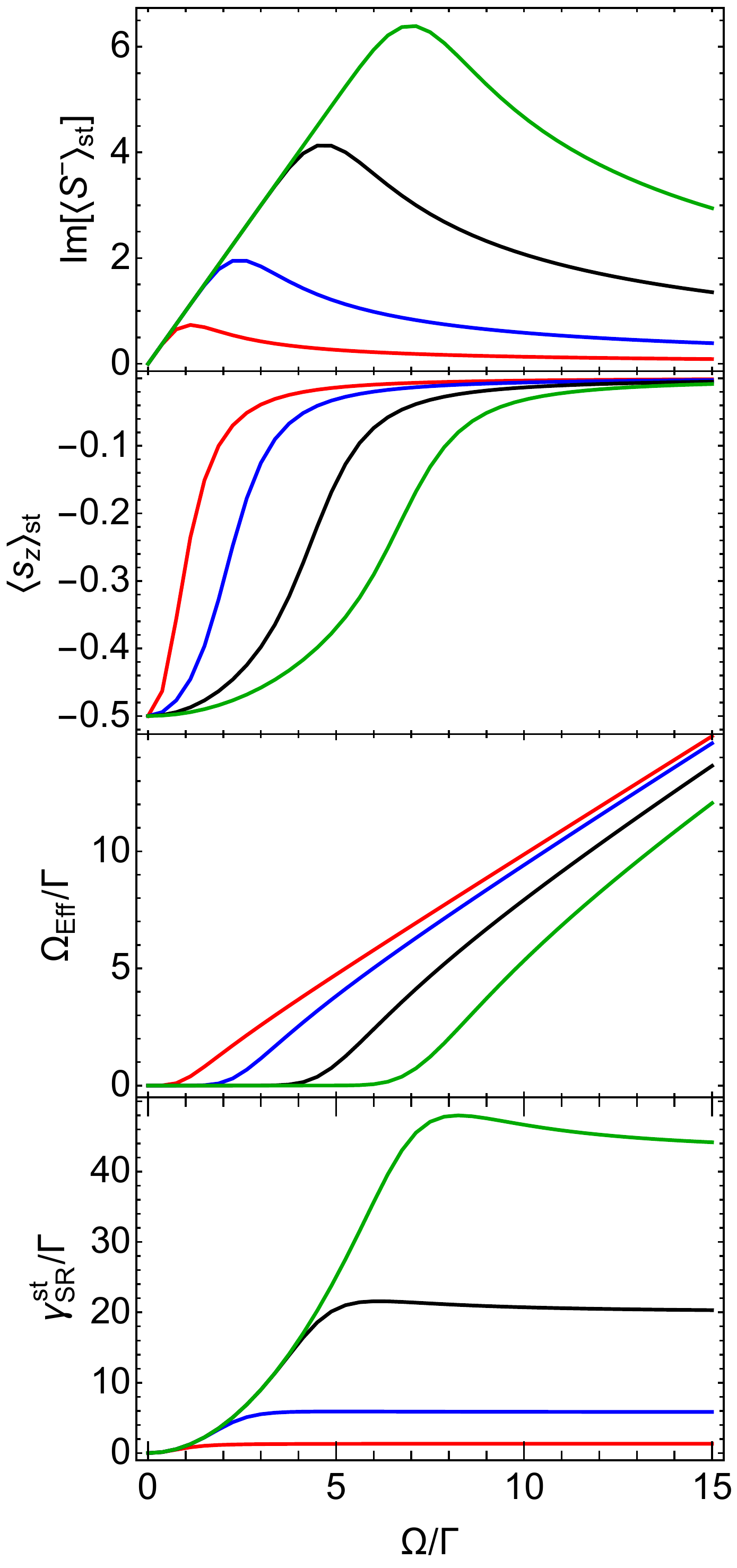}
	\caption{\textbf{Steady state values} of the collective dipole
	${\rm Im}[\langle S^-\rangle]$, the magnetiztion $\langle s_z\rangle$, the effective Rabi frequency
	$\Omega_{\rm Eff}$ and the superradiant emission rate $\gamma_{\rm SR}$ as a function of $\Omega/\Gamma$, 
	plotted for $N=(2,5,10,15)$ (red, blue, black, green).}
	\label{figsteadystatetheo}
\end{figure}

\subsubsection{ Numerical solutions of the DDM}\label{App:PredictionDDM_numerical}
Figure~\ref{figsteadystatetheo} shows the results for the collective dipole ${\rm Im}[\langle S^-\rangle]$, 
the magnetiztion $\langle s_z\rangle$, the effective Rabi frequency 
$\Omega_{\rm Eff}=\Omega - i\Gamma\langle S^-\rangle$ and the superradiant emission rate 
$\gamma_{\rm SR}$ as a function of the excitation laser Rabi frequency $\Omega$. We observe two regimes.  

In the first one, corresponding to  $\Omega/\Gamma \leq N/2$, 
the collective dipole ${\rm Im}[\langle S^-\rangle]$ is proportional to $\Omega$ and the screening of the
driving field by the field scattered by the collective dipole is efficient. 
To better understand the screening, we consider the 
limiting case where  $\Omega/\Gamma \ll N/2$.  We may then restrict ourselves to the two lowest Dicke states, 
$|N/2,-N/2\rangle$ and $|N/2,-N/2+1\rangle$, corresponding respectively to $| G\rangle=|ggg...g\rangle$ 
and $|W\rangle=(|egg...g\rangle+|geg...g\rangle+...|ggg...e\rangle)/\sqrt{N}$. 
The matrix element of the collective dipole connecting the two states is $d\sqrt{N}$, with $d$ the single-atom dipole, 
so that the decay rate of $|W\rangle$ is $N\Gamma$ and  the collective coupling to the laser is $\Omega\sqrt{N}$.
Restricting ourselves to this two-level system, $\gamma_{SR}= N\Gamma \pi_W$ where 
$\pi_W \approx (\Omega\sqrt{N})^2/(N\Gamma)^2 $ is the population of the $|W\rangle$ state. Hence, 
$\gamma_{SR}= \Omega^2/\Gamma$, independent of $N$.   
Similarly, $\langle S^-\rangle= i\sqrt{N}(\Omega\sqrt{N})/(N\Gamma)$, also independent of $N$.  
As we approach $\Omega/\Gamma \lesssim N/2$, however, $\langle S^-\rangle$ remains proportional
to $\Omega$ despite the fact that we significantly populate the Dicke states up to 
$m\approx0$: the suppression of the coherences $\rho_{m,m-1}$ due to the strong driving is counteracted by the enhanced 
coupling $A_{m-1}$ between Dicke states $|S,m\rangle$ and $|S,m-1\rangle$. 
Thus, despite the saturation of the lowest Dicke states,
a collective dipole corresponding to a collective Bloch vector can develop even close to the equatorial plan. 
This would be impossible for a two-level system and this is a feature of the ladder of Dicke states. 

In the second regime, $\Omega/\Gamma \gg N/2$, the system is saturated, 
the collective dipole $\langle S^-\rangle\rightarrow 0$, 
and the population of  each Dicke state $|S,m\rangle$ is $\rho_{m,m}=1/(N+1)$. 
Calculating the sum in Eqs.~(\ref{Eq:spinvariabletheo}), we get $\gamma_{\rm SR} = N(N+2)/6$, independent of $\Omega$. 

\begin{figure}
	\includegraphics[width=0.9\linewidth]{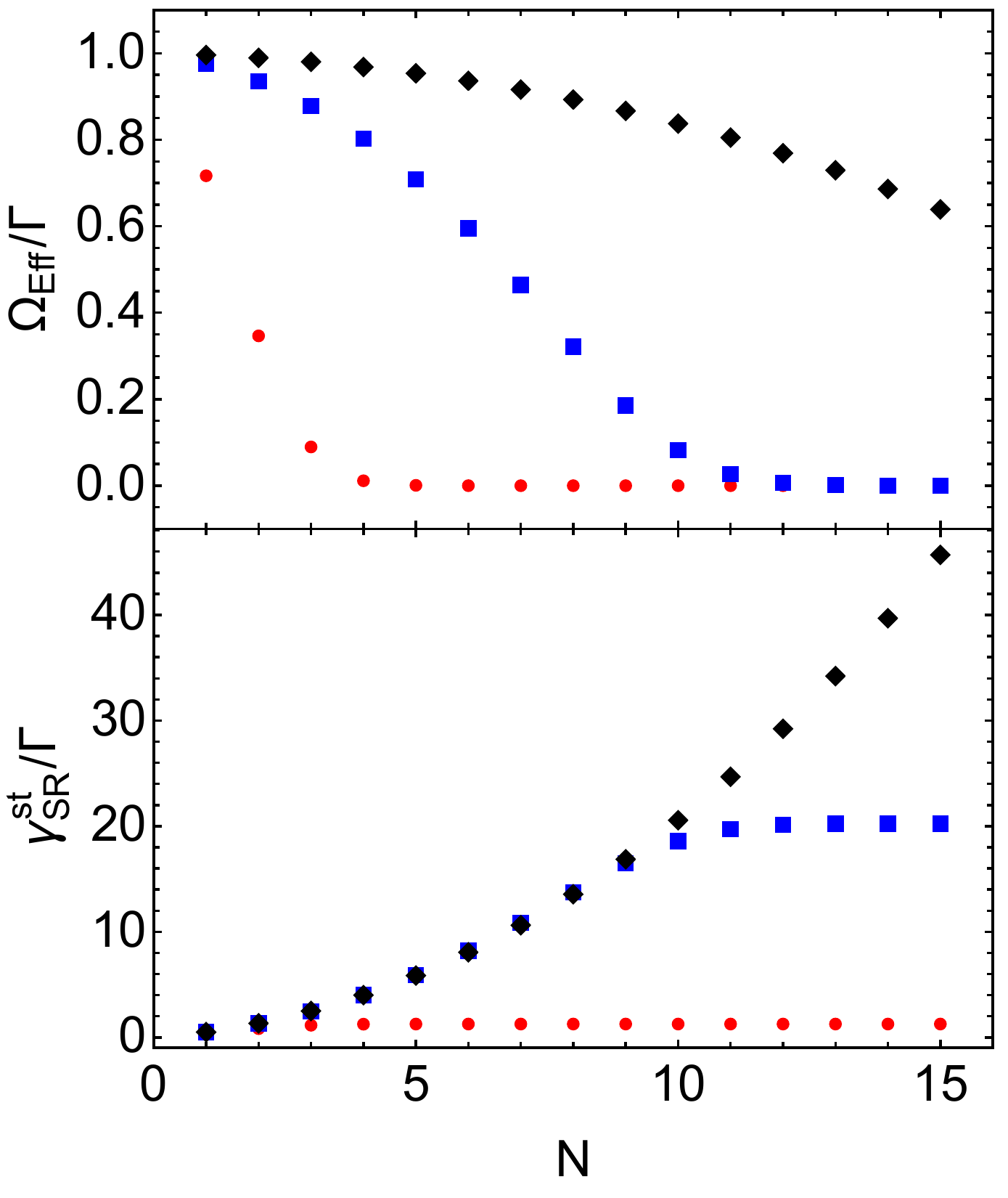}
	\caption{\textbf{Steady state values} of $\Omega_{\rm Eff}$, and $\gamma_{\rm SR}$  as a function of $N$, for 
	$\Omega/\Gamma = (1.1,4.5,11)$ (red dots, blue squares, black diamonds).	}
	\label{figsteadystatetheo_N}
\end{figure}

We also plot in Fig.~\ref{figsteadystatetheo_N} $\Omega_{\rm Eff}$ and $\gamma_{\rm SR}$
as a function of $N$ for different $\Omega$. We confirm that for $N\ge 2\Omega/\Gamma$
the screening from the collective dipole operates, and that $\gamma_{SR}\rightarrow \Omega^2/\Gamma$
for $N\rightarrow\infty$. 
We also observe that for increasing values of $N$ starting from 1, 
$\gamma_{SR}\propto N^\alpha$, with $\alpha$ decreasing  from $2$ to 0. 
Such a decrease is observed in the experiment (Fig.~\ref{fig_steady}(c)).

\subsubsection{Analytical derivation of the phase transition}\label{App:PT}
Finally, we briefly rederive the
prediction of a second-order phase transition in the thermodynamics limit ($N\rightarrow \infty$), 
following Ref.~\cite{Walls_1980}. The field operator $\hat E_{\rm Eff}^+$ inside the cloud is the superposition 
of the classical laser field $\hat E_{\rm L}^+$ and of the fields scattered by all the atoms, $\hat E_{\rm sc}^+=-i\hbar\Gamma  \hat S^-/d$.
Hence, $\hat E_{\rm Eff}^+=\hat E_{\rm L}^+-i\hbar\Gamma  \hat S^-/d$, leading to
\begin{equation}
\Omega^2={\hbar ^2\over d^2}\langle \hat E_{\rm L}^-E_{\rm L}^+\rangle\approx
\Omega_{\rm Eff}^2+{N^2\Gamma^2\over 4}\langle \hat\sigma^+\hat\sigma^-\rangle\ ,
\end{equation}
with $\hat S^-=N \hat\sigma^-/2$ and $\Omega_{\rm Eff}^2=\hbar ^2\langle \hat E_{\rm Eff}^-E_{\rm Eff}^+\rangle/d^2$. 
We have neglected here the terms $\langle E_{\rm Eff}^- \hat S^+\rangle$ 
and $\langle E_{\rm Eff}^+ \hat S^-\rangle$, which are of order $N$ only. Each two-level atom in the cloud is driven by the 
effective Rabi frequency $\Omega_{\rm Eff}$, hence:
\begin{equation}
\langle \hat\sigma^+\hat\sigma^-\rangle={1\over 2}{2\Omega_{\rm Eff}^2/\Gamma^2\over1+ 2\Omega_{\rm Eff}^2/\Gamma^2}
\end {equation}
in steady-state and on resonance ($\Delta = 0$)~\cite{AllenEberly}.
Introducing $\beta=2\Omega/(N\Gamma)$ and $x=2\Omega_{\rm Eff}/(N\Gamma)$ yields:
\begin{equation}\label{Eq:xsol}
\beta^2= x^2+ {N^2x^2/2\over 1+ N^2x^2/2}\ .
\end{equation}
Considering that $N\gg 1$, $x^2\approx \beta^2-1$ for $\beta \ge 1$, hence 
\begin{equation}\label{Eq:xcritical}
\beta\ge 1\Rightarrow x\approx \sqrt{\beta^2-1}\ .
\end{equation}
For $\beta < 1$, we get $x\ll 1$, so that:
\begin{equation}\label{Eq:scalingx}
\beta^2\approx{N^2x^2/2\over 1+ N^2x^2/2} \Rightarrow x\approx {\sqrt{2}\over N}{\beta\over \sqrt{1-\beta^2}}\ .
\end{equation} 
These two last equations show the existence of a critical point for $\beta = 1$, with $\Omega_{\rm Eff}=0$ for $\beta <1$ 
when $N\rightarrow \infty$, reminiscent of a second order phase transition. 

\begin{figure}
	\includegraphics[width=0.9\linewidth]{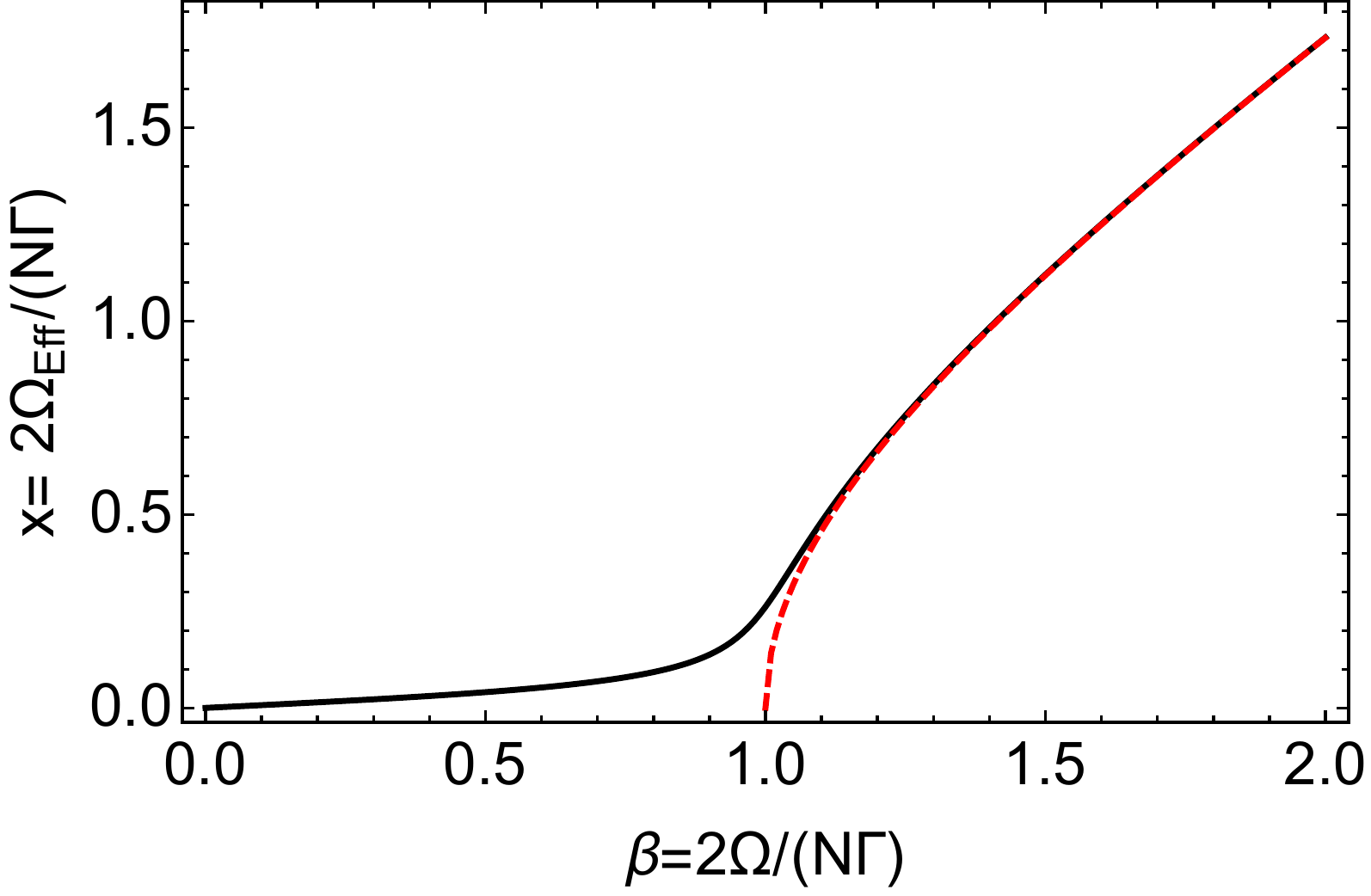}
	\caption{\textbf{DDM and second order phase transition. } Comparison between the numerical solution of Eq.\,(\ref{Eq:xsol}) for 
	$N=20$ (black line), and the analytical solution $x= \sqrt{\beta^2-1}$ (red dashed line), 
	showing the existence of a critical point for $N\rightarrow \infty$.}
	\label{Fig_analyticalDDM}
\end{figure}

It may look inconsistent to obtain $\Omega_{\rm eff}= 0$ while $\langle \hat S^-\rangle =-i\Omega/\Gamma\neq 0$
in the case $\beta <1$, as the effective field is the source of $\hat S^-$. 
However, for large but finite $N$, $\Omega_{\rm Eff}={\cal O}(1/N)$ [Eq.(\ref{Eq:scalingx})], 
so that  $\langle \hat S^-\rangle \propto N \Omega_{\rm Eff}$ remains finite. 
Figure~\ref{Fig_analyticalDDM} shows the numerical solution of 
Eq.\,(\ref{Eq:xsol}) for $N=20$, together with the analytical solution  of Eq.\,(\ref{Eq:xcritical}). 

\subsection{Cooperative coupling between an atomic ensemble and a diffraction mode}\label{App:Cooperativity}

In this section, we evaluate the cooperative coupling  between a 
generic atomic distribution and a diffraction mode in free space.
The intensity emitted in a direction $\bf k$ by a cloud containing $N$ atoms is given by \cite{AllenEberly}:
\begin{equation}
	I_N({\bf k})= I_1({\bf k})\big{[} \sum_i^N {\langle \hat{\sigma}^z_i \rangle+1\over 2} + \sum_{i\neq j}^N e^{i {\bf k}\cdot({\bf r}_i-{\bf r}_j)} \langle \hat{\sigma_i^+}\hat{\sigma_j^-} \rangle \big{]}
\label{Eq:in}
\end{equation}
where $I_1({\bf k}) (\langle \hat{\sigma}^z_i \rangle+1)/2$ is the single atom intensity. 
The first term on the right side of Eq.\,\eqref{Eq:in} 
is the {\it incoherent} intensity emitted by the system. 
The second term describes the correlations between different atoms and 
is responsible for the {\it coherent} part of the emission. 

As done in \cite{AllenEberly} for the non-driven case,  
we now assume a collective, factorizable atomic state of the $N$ atoms, 
excited by a laser with a wavevector ${\bf k}_{\rm L}$: 
$\langle \hat{\sigma_i^+}\hat{\sigma_j^-} \rangle\approx \langle \hat{\sigma_i^+}\rangle\langle\hat{\sigma_j^-} \rangle
=|\langle \hat{\sigma^+}\rangle|^2\, e^{-i {\bf k}_{\rm L}\cdot({\bf r}_i-{\bf r}_j)}$.
The coherent part of the radiation is then:
\begin{equation}
	I_N^{\rm coh}({\bf k})=I_1({\bf k})|\langle \hat{\sigma^+}\rangle|^2\sum_{i\neq j}^N e^{i({\bf k}-{\bf k}_{\rm L})\cdot({\bf r}_i-{\bf r}_j)}\ .
	\label{Eq:inC}
\end{equation}
Introducing the structure factor:
\begin{equation}
 	\Gamma({\bf k},{\bf k}_{\rm L})= \frac{1}{N^2}\sum_{i\neq j}^N e^{i({\bf k}-{\bf k}_{\rm L})\cdot({\bf r}_i-{\bf r}_j)}
	\label{Eq:gamma}
\end{equation}
leads to:
\begin{equation}
	I_N^{\rm coh}({\bf k} ) = N^2 I_1({\bf k})|\langle \hat{\sigma^+}\rangle|^2 \Gamma({\bf k},{\bf k}_{\rm L})\ .
	\label{Eq:inD}
\end{equation}
In analogy with cavity QED, we define the effective atom number coupled to 
the diffraction mode (extending over a solid angle $\Delta\Theta$) as
$N\mu =P_N^{\rm coh}/(NP_1)$
where $P_N^{\rm coh}$ is the power radiated by $N$ atoms into the diffraction mode, i.e., 
$P_{N}^{\rm coh}= \int_{4\pi} d\Omega_{{\bf k}}I_{N}({\bf k})$ and $P_1$ 
is the power radiated by a single atom in $4\pi$. 
In the weak driving regime, $|\langle \hat{\sigma^+}\rangle|^2\approx (\langle \hat{\sigma}^z_i \rangle+1)/2$ and we get:
\begin{equation}
  	N\mu=\frac{N^2\displaystyle\int_{4\pi} d\Omega_{{\bf k}}I_1({\bf k}) 
	\Gamma ({\bf k}, {\bf k}_{\rm L})}{N\displaystyle\int_{4\pi} d\Omega_{{\bf k}}I_1({\bf k})} \ .
	\label{Eq:coop2}
\end{equation} 
As the structure factor $\Gamma({\bf k}, {\bf k}_{\rm L})$ 
has non-zero values only in $\Delta\Theta$, $\mu\sim \Delta \Theta /(4\pi)$.
This derivation shows that the effective atom number comes from the
shape factor introduced in the context of the (non-driven) superradiance in extended 
clouds \cite{AllenEberly,superradiance2017sutherland}. Note that the value of 
$\mu$ depends on the direction of the excitation laser.

We now calculate $\mu$ for the specific geometry of our experiment. 
For circularly polarized dipoles, $I_1({\bf k})=I_1(\phi, \theta)=(1+\cos^2\phi\sin^2\theta)/2$, with $\theta$ 
and $\phi$ the polar and azimuthal angles with respect to the quantization axis.  
($\hat{k}= (\cos{\theta}, \sin{\theta}\cos{\phi},\sin{\theta}\sin{\phi})$). 
Then, $P_1\propto 8\pi/3$. 
To calculate  $\Gamma({\bf k}, {\bf k}_{\rm L})$ and $P_N^{\rm coh}$, 
we replace the sum over discrete positions by an integral over a continuous density distribution $\rho({\bf r})$: 
\begin{equation}
	\Gamma({\bf k}, {\bf k}_{\rm L})=\left|\int d^3{\bf r}\, \rho ({\bf r})\, e^{i ({\bf k}-{\bf k}_{\rm L})\cdot {\bf r}}\right|^2\ .
	\label{Eq:gamma1}
\end{equation}
Assuming a Gaussian density $\rho ({\bf r})$ with r.m.s. size $\ell_{\text{ax}}$ along $\hat{x}$ and 
$\ell_{\text{rad}}$ in $\hat{y}$ and $\hat{z}$, setting ${\bf k}_{\rm L}$ along $\hat x$, we obtain: 
\begin{align}
	P_N^{\rm coh}&\propto \pi \int_0^\pi d\theta \sin\theta \big{(}1+\frac{\sin^2\theta}{2}\big{)}\times \nonumber\\
	&\exp[-(k\ell_{\text{rad}}\sin{\theta})^2]\, \exp[-(k\ell_{\text{ax}})^2(\cos{\theta}-1)^2]\ .
\end{align}
A Taylor expansion of the second exponential in the integral indicates that the integrant is 
non negligible in the solid angle $\Delta\Theta/(4\pi)\sim\mu \sim \lambda/(2\pi \ell_{\rm ax})$.
With the experimental values $\ell_{\text{ax}}$ and $\ell_{\text{rad}}$, we get 
$\mu \simeq 2.5 \times 10^{-3}$, a factor 2 smaller than the value used in the main text, 
obtained as a free parameter. However, a precise estimation of the trap size is challenging and 
subjected to overestimation (due to radiation pressure effects for instance). 
Considering a $50\%$ error in the measure of $\ell_{\text{ax}}$ makes the result 
consistent with the value used in the main text. 

Importantly, the structure factor, and consequently the value of $\mu$, depends both on the
cloud geometry and the direction of the excitation laser  ${\bf k}_{\rm L}$. 
In previous works \cite{ferioli2021laser, Glicenstein:22}, we excited the same atomic ensemble,  but
perpendicularly to the main axis. This led to a value of $\mu$ smaller than  
the one achieved here by one order of magnitude. 
Consequently, in that case, $\tilde{N}\ll 1$, making it impossible to observe the phase transition. 

{\color{black}
Finally, we compare the magnitudes of the two contributions in Eq.~(\ref{Eq:in}). 
Assuming that the correlations between atomic dipoles are perfect 
($\langle \hat{\sigma_i^+}\hat{\sigma_j^-} \rangle\sim 1$) and that the atoms are nearly saturated
($\langle \hat{\sigma_i^z} \rangle\approx 0$), the contribution of the correlation terms in the transverse
direction is $\sim N\, \Gamma({\bf k},{\bf k}_{\rm L})$ smaller than the population term. For our experimental configuration 
$\Gamma({\bf k},{\bf k}_{\rm L})\sim \exp[-(k\ell_{\text{ax}})^2]\sim \exp(-120^2)$, {i.e.} totally negligible. 
This indicates that the dominant contribution 
to the APD$\perp$-signal does come from the atomic population. }

\end{document}